%% file: RR-9162.tex
\documentclass[twoside]{article}
\input{revision_macros.tex}
\usepackage[a4paper]{geometry}
\usepackage[T1]{fontenc} % ou \usepackage[OT1]{fontenc}
\usepackage{RR}

\usepackage{algorithm}
\usepackage{algorithmicx}
\usepackage[noend]{algpseudocode}
\usepackage{amsmath}
\usepackage{amssymb}
\usepackage{float}
\usepackage{graphicx}
\usepackage[utf8]{inputenc}
\usepackage{inputenc} % Include this package again for the i of Taiani
\usepackage{subcaption}
\usepackage{url}
\usepackage{xspace}
\usepackage{verbatim}
\usepackage{wrapfig} % Gives figures borders

\graphicspath{ {./}{figs/} }

\newcommand{\figref}[1]{Figure~\ref{#1}}
\newcommand{\secref}[1]{Section~\ref{#1}}
\newcommand{\algoref}[1]{Algorithm~\ref{#1}}

\newcommand*\circledblack[1]{\tikz[baseline=(char.base)]{
            \node[shape=circle,draw,inner sep=0.5pt,fill,text=white] (char) {\sf \small #1};}}

\renewcommand{\change}[4]{#1}
\renewcommand{\annote}[3]{}

\RRNo{9162}
\RRdate{March 2018}
\RRauthor{% les auteurs
		Davide Frey\thanks[UnivR]{Univ Rennes, Inria, CNRS, IRISA, France},
		Marc X. Makkes\thanks{Vrije Universiteit Amsterdam, The Netherlands},
		Pierre-Louis Roman\thanksref{UnivR},\\
		François Taïani\thanksref{UnivR},
		Spyros Voulgaris\thanks{Athens University of Economics and Business, Greece}
    }
\authorhead{Frey \& Makkes \& Roman \& Taïani \& Voulgaris}
\RRtitle{Dietcoin: court-circuiter la vérification dans Bitcoin pour les téléphones mobiles}
\RRetitle{Dietcoin: shortcutting the Bitcoin verification process for your smartphone}
\titlehead{Dietcoin}
\RRresume{\input{dietcoin_resume_fr}}
\RRabstract{
Blockchains have a storage scalability issue.
Their size is not bounded and they grow indefinitely as time passes.
As of August 2017, the Bitcoin blockchain is about 120\,GiB big while it was only 75\,GiB in August 2016.
To benefit from Bitcoin full security model, a bootstrapping node has to download and verify the entirety of the 120\,GiB.
This poses a challenge for low-resource devices such as smartphones.
Thankfully, an alternative exists for such devices which consists of downloading and verifying just the header of each block.
This partial block verification enables devices to reduce their bandwidth requirements from 120\,GiB to 35\,MiB.

However, this drastic decrease comes with a safety cost implied by a partial block verification.
In this work, we enable low-resource devices to fully verify subchains of blocks without having to pay the onerous price of a full chain download and verification; a few additional MiB of bandwidth suffice.
To do so, we propose the design of diet nodes that can \emph{securely} query full nodes for shards of the UTXO set, which is needed to perform full block verification and can otherwise only be built by sequentially parsing the chain.
 }
\RRmotcle{blockchain, partitionnement, UTXO, registre distribué, monnaie cryptographique, informatique mobile}
\RRkeyword{blockchain, sharding, UTXO, distributed ledger, cryptocurrency, mobile computing}
\RRprojets{WIDE}
\RCRennes % Rennes - Bretagne Atlantique
\begin{document}
\makeRR

\tableofcontents

\input{algos/algo_include}

% ===== 12 pages double columns
% 1.5 pages: abstract + intro
% 0.5 pages: problem statement
% 2   pages: background on bitcoin
% 2   pages: dietcoin system
% 2   pages: security analysis
% 2.5 pages: performances evaluation
% 1   pages: related work
% 0.5 pages: conclusion + funds
% x   pages: bib + annexes

\input{dietcoin_introduction}

\input{dietcoin_background}
\input{dietcoin_system}
%\input{dietcoin_analysis}
%\input{dietcoin_evaluation}
\input{dietcoin_related}

\input{dietcoin_conclusion}

\section*{Acknowledgments}
This work has been partially funded by the Region of Brittany, France,
by the Doctoral school of the University of Brittany Loire (UBL), by
the French National Research Agency (ANR) project \emph{SocioPlug} under
contract ANR-13-INFR-0003 (\url{http://socioplug.univ-nantes.fr}) and by
the SIDN Fonds contract 172027.

\bibliographystyle{IEEEtran}
\bibliography{dietcoin}

\newpage
\input{algos/algo_tables.tex}
\input{algos/algo_spv_compact.tex}

\end{document}

%% file: revision_macros.tex
\usepackage{ifthen}
\usepackage{calc}
\usepackage[usenames,dvipsnames,pdftex]{color}
\usepackage[normalem]{ulem}
\usepackage{tikz}

\newcommand{\annote}[3]{{
		\colorbox{#3}{\bfseries\sffamily\footnotesize\textcolor{white}{#2}}
		\color{#3}
		$\blacktriangleright${\it #1}$\blacktriangleleft$}
}

\newcommand{\commentPLR}[1]{\annote{#1}{PLR}{blue}}

\DeclareRobustCommand{\change}[4]{{\color{#3}#1}\ifx\relax#4\relax\else\xspace\textcolor{Gray}{\sout{#4}}\fi}

%% file: dietcoin_resume_fr.tex
% -*- tex-main-file: "RR-9162" compile-command: "make pdf" ispell-dictionary: "francais" -*-

Les blockchains telles que Bitcoin passent mal à l'échelle, notamment du fait de leurs besoins important de stockage.
Les besoins de stockage d'une blockchain typique ne sont pas limités et croissent indéfiniment.
Par exemple, en août 2017, les données contenues dans la blockchain Bitcoin représentaient environ 120 \,GiB, contre 75\,GiB un an auparavant, en août 2016.
Pour bénéficier des garanties complètes de sécurité apportées par Bitcoin, un n\oe{}ud qui rejoint le réseau doit télécharger et vérifier l'intégralité des 120 \,GiB de données.
Cette nécessité pose un défi pour les appareils à faibles ressources tels que les smartphones.
Heureusement, une alternative existe pour de tels dispositifs qui consiste à télécharger et à vérifier seulement l'en-tête de chaque bloc de la blockchain.
Cette vérification partielle des blocs permet aux appareils de réduire leurs besoins en bande passante de 120 \,GiB à 35 \,MiB, mais cette diminution drastique ne permet qu'une vérification partielle des blocs, et diminue grandement les garanties de sécurité offertes aux n\oe{}uds qui l'utilisent.

Dans ce travail, nous proposons une approche qui permet aux appareils à faibles ressources de vérifier entièrement des sous-chaînes de blocs sans avoir à payer le prix onéreux d'un téléchargement et d'une vérification complète de la chaîne ; quelques MiB supplémentaires de bande passante suffisent.
Pour ce faire, nous proposons d'introduire des n\oe{}uds Dietcoin qui sont capables \emph{en toute sécurité} d'interroger des n\oe{}uds exécutant le protocole complet pour obtenir des fragments d'un ensemble appelé UTXO, nécessaire à la vérification complète des blocs.

%% file: algos/algo_include.tex
% -*- tex-main-file: "dietcoin" compile-command: "make pdf" ispell-dictionary: "american" -*

\algblockdefx[Event]{Event}{EndEvent}[1]{\textbf{upon event} (#1) \textbf{do}}{}
\algblockdefx[Init]{Init}{EndInit}{\textbf{initialization}}{}
\algblockdefx[Parameters]{Parameters}{EndParameters}{\textbf{parameters}}{}
\algblockdefx[Periodically]{Periodically}{EndPeriodically}{\textbf{periodically}}{}
\algblockdefx[Receive]{Receive}{EndReceive}[1]{\textbf{upon receive} (#1) \textbf{do}}{}

\makeatletter
\ifthenelse{\equal{\ALG@noend}{t}}
	{\algtext*{EndEvent}}
	{}
\ifthenelse{\equal{\ALG@noend}{t}}
	{\algtext*{EndInit}}
	{}
\ifthenelse{\equal{\ALG@noend}{t}}
	{\algtext*{EndParameters}}
	{}
\ifthenelse{\equal{\ALG@noend}{t}}
	{\algtext*{EndPeriodically}}
	{}
\ifthenelse{\equal{\ALG@noend}{t}}
	{\algtext*{EndReceive}}
	{}
\makeatother

% Comments in italic
\algrenewcommand{\algorithmiccomment}[1]{\hfill \(\triangleright\) {\textit{#1}}}
\algnewcommand{\LeftComment}[1]{\State \(\triangleright\) \textit{#1}}

% Highlight dietcoin algo with colored bullets next to line number (\bc bitcoin, \dc dietcoin)
\newcommand{\bc}{\algrenewcommand{\alglinenumber}[1]{\footnotesize $\circ$##1:}}
\newcommand{\dc}{\algrenewcommand{\alglinenumber}[1]{\footnotesize $\bullet$##1:}}

% \newcommand{\bc}{ }
% \newcommand{\dc}{ }

% Or use adjustbox? Or just space right the line numbers
% https://tex.stackexchange.com/questions/64674/coloring-lines-in-an-algorithm/64714#64714
%\usepackage{xcolor}
%\usepackage{adjustbox}

% Keywords
\newcommand{\kw}[1]{\textbf{#1}\xspace}
% Functions
\newcommand{\fn}[1]{\textsc{#1}}
% Global variables
\newcommand{\gv}[1]{\mathsf{#1}_p}
% Local variables
\newcommand{\lv}[1]{\mathsf{#1}}

% Predefined functions
\newcommand{\bloomFilter}{\mathrm{bloomFilter}}
\newcommand{\depth}{\mathrm{depth}}
\newcommand{\difficulty}{\mathrm{difficulty}}
\newcommand{\getShardKey}{\mathrm{getShardKey}}
\newcommand{\getShardingPolicy}{\mathrm{getShardingPolicy}}
\newcommand{\hashFunction}{\mathrm{HASH}}
\newcommand{\height}{\mathrm{height}}
\newcommand{\initTemplate}{\mathrm{initTemplate}}
\newcommand{\buildMRoot}{\mathrm{buildMRoot}}
\newcommand{\sha}[1]{\textsc{SHA-256(SHA-256}(#1))}
\newcommand{\updateBestTip}{\mathrm{updateBestTip}}
\newcommand{\updateDifficulty}{\mathrm{updateDifficulty}}
\newcommand{\updateMTreeInPlace}{\mathrm{updateMTreeInPlace}}

%%% Local Variables:
%%% mode: latex
%%% End:

%% file: dietcoin_introduction.tex
% -*- tex-main-file: "RR-9162" compile-command: "make pdf" ispell-dictionary: "american" -*

\section{Trustless Bitcoin}
\label{sec:introduction}

Within a decade, blockchains have become extremely popular, and have
been used to implement several widely-used
crytocurrencies~\cite{nakamoto_bitcoin_2008}, and smart-contract
services~\cite{DBLP:conf/podc/DickersonGHK17a}. A blockchain implements a \emph{tamper-proof
  distributed ledger} in which public transactions can be recorded in
a close-to-irrevocable manner. Recorded transactions are stored into
\emph{blocks}, which are then incrementally linked (or \emph{chained})
in order to form an append-only list. The irrevocability of these
chaining mechanisms exploits cryptographic mechanisms and peer-to-peer
exchanges. This combination makes it in principle inconceivably
hard for individual participants to revoke past transactions
(due to the computational cost involved), while it remains possible for any
participant to verify the validity of a blockchain's entire history.

%% Bitcoin~\cite{nakamoto_bitcoin_2008} is an innovative payment system designed to function without any central authority.
%% A central authority typically ensures the well processing of data and certifies the correctness of the content it decides upon.
%% Users of a system can trust the system if they trust the central authority.

%% However, how can users trust a system lacking such an authority?
%% Blockchains answer that question by making every data verifiable and let users verify the validity of data by themselves, for themselves.
%% With blockchains, users do not need to trust any other actor and they can trust the system as long as they trust the software and hardware that runs it.

\begin{figure}
	\centering
	\includegraphics[width=0.5\textwidth]{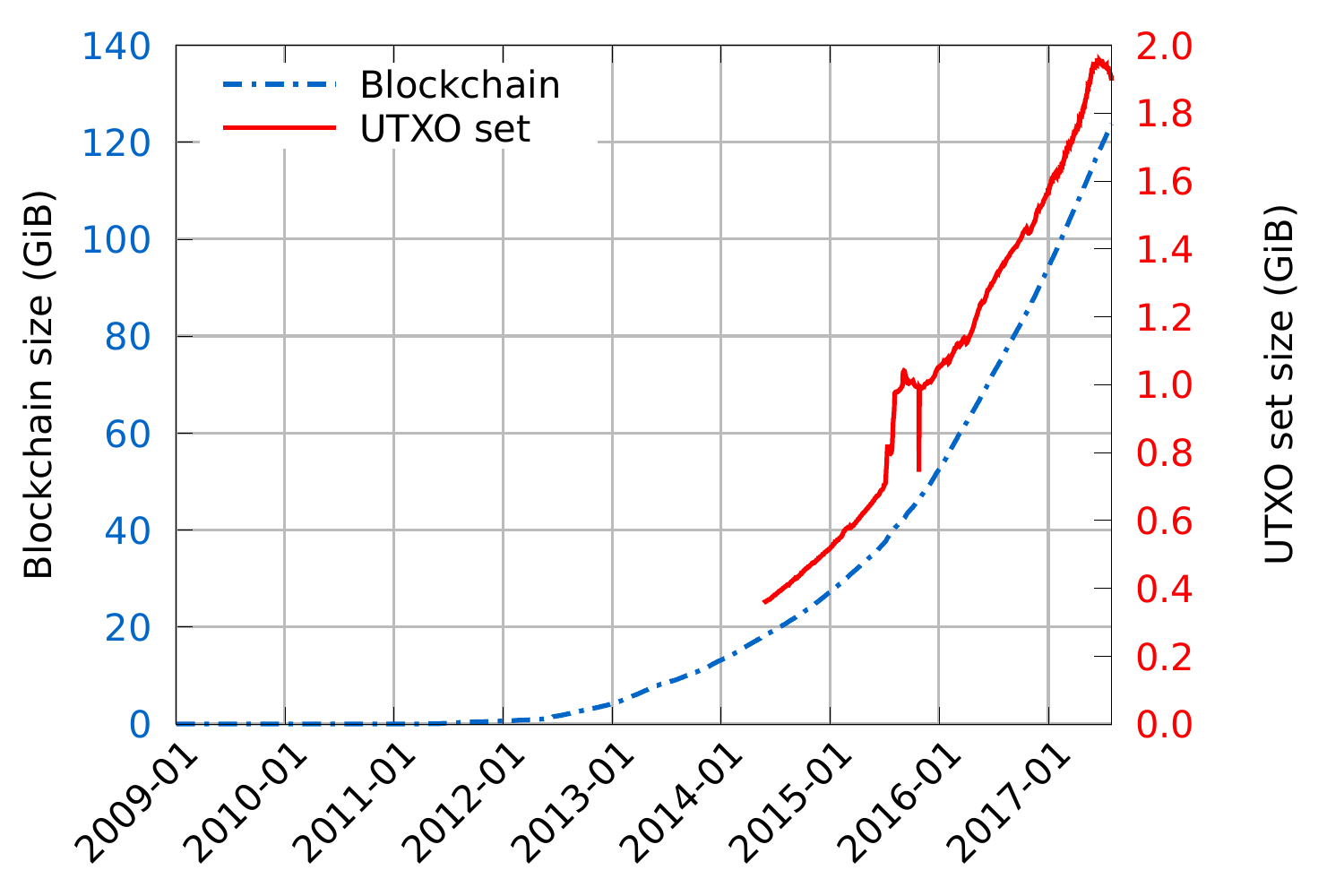}
	\caption{Both the Bitcoin blockchain and the UTXO set have almost tripled in size in the past two years.}
  
	\label{fig:sizes}
\end{figure}

Verifying a blockchain remains, however, a particularly costly
process. The verifying node must first download the entire blockchain,
which in many cases has reached a size beyond the communication
capabilities of many mobile
devices. %% . Because a blockchain grows constantly as new transaction are appended to it, current blockchains have reached sizes that make it hard for constrained devices to fully participate in the blockchain's protocol.
The Bitcoin blockchain, for instance, had grown to 120\,GiB as of
August 2017~(\figref{fig:sizes}), and follows an exponential growth,
implying the problem can only become more acute.

Once the blockchain has been downloaded, the verifying node must then
check its consistency block by block, a lengthy process that can take
hours on high-end machines.
% that are intractable for constrained devices, such as mobile phones
%% be it in terms of communication, storage, or computation.
The exorbitant price of a full chain verification makes it unrealistic for low-resource devices to fully implement a blockchain protocol.
Some blockchain systems, such as Bitcoin, therefore enable nodes to perform varying degrees of verification: \emph{full nodes} verify everything while lightweight nodes only verify a small fraction of the data.

%% To benefit from the full security that Bitcoin has to offer, a user can chose to deploy a full node that verifies every element composing the chain.
%% To do so, the chain must first be downloaded, which represents a bandwidth cost of 120\,GiB as of August 2017 and which cost rapidly grows as shown in \figref{fig:sizes}.

%% The chain must then be entirely verified block by block in a lengthy process that can take hours on high end machines.
%% The vast majority of the computation time is spent on verifying transactions, ensuring that the money spent is indeed spendable and is spent by its rightful owner.

In the case of Bitcoin, this lightweight verification is known as \emph{Simplified Payment Verification} (SPV for short). SPV nodes only download and verify a much reduced version of the Bitcoin blockchain, comprised only of its block headers, which today only weights 35\,MiB (a reduction by three orders of magnitude). This summary version however only contains the chaining information making up the blockchain, not the recorded transactions. This information is sufficient for SPV nodes to verify that the chain's structure is valid (and hence very unlikely to have been created by malicious nodes), but not that a past transaction does exist in the chain.
As a result, SPV nodes are vulnerable to attacks in which an attacker leads an SPV node to believe a transaction $t$ has occurred, while $t$ is later on rejected by the system because the funds transferred by $t$ have in fact already been spent (known as a \emph{double-spend attack}).

To protect themselves against double-spend attacks, full nodes keep track of unspent funds %% by caching them into
in a structure known as the set of \emph{Unspent TransaCTion Outputs} (UTXO set). The UTXO set is unfortunately costly to construct
(as this construction requires the entire blockchain), to exchange (currently weighing 1.9\,GiB, see \figref{fig:sizes}), and to maintain, which explains why SPV nodes do not use it.

%% while full nodes are not, proving a trade-off between security and bandwidth/time requirement for nodes.

%% The vast majority of the computation time is spent on verifying the validity of transactions, i.e. in the case of Bitcoin ensuring that the money spent is indeed spendable and is spent by its rightful owner.

%% To help in this task, full nodes keep track of the spendable coins at all time by caching them into the UTXO set (Unspent TransaCTion Output) to ensure faster verification time.

%% A much more lighweight strategy consists in deploying an SPV node (Simplified Verification Payment) that only downloads and verifies the 35\,MiB of block headers from the chain.
%% In effect SPV nodes do not verify the correctness of the transactions contained in blocks, they rather rely on full nodes to perform this task.
%% Even though a header only verification shields against most cheap attacks, such a drastic decrease in cost has security repercussions.
%% It is perfectly possible for an attacker to craft a block with a valid header and yet with invalid transactions.
% There is a non-null probability that a block $b$ with a valid header contains fake txs, this probability decreases (exponentially?) as new blocks are appended to the chain containing $b$. Since blocks with a small depth are easier to fake, we propose to increase the trust of Diet nodes in them by fully verifying these blocks instead of just verifying their header difficulty.

In this report, we propose to bridge the gap between full nodes and SPV nodes by introducing \emph{diet nodes}, and their associated protocol, \emph{Dietcoin}. % derived from SPV nodes.
Dietcoin strengthens the security guarantees of SPV nodes by bringing them close to those of full nodes. Dietcoin enables low-resource nodes to verify the transactions contained in a block without constructing a full-fledged UTXO set. 
In our protocol, diet nodes download from full nodes only the parts of the UTXO set they need in order to verify a transaction of interest. This selective download mechanism must, however, be realized with care.
Diet nodes must be able to detect any tampering of the UTXO set itself,  at a cost that remains affordable for low-resource devices, both in terms of communication and computing overhead.

The rest of this report is structured as follows. We first present the Bitcoin protocol in more detail (\secref{sec:bitcoin}), and explain the workings of full and SPV nodes. We then detail the design of Dietcoin and diet nodes and discuss the security guarantees they provide (\secref{sec:dietcoin}).
Finally, we present related work (\secref{sec:related_work}), and conclude (\secref{sec:conclusions}).

%% file: dietcoin_background.tex
\section{The Bitcoin system}
\label{sec:bitcoin}

%\begin{figure*}[!t]
%	\begin{minipage}[b]{0.49\textwidth}
%		\centering
%		\includegraphics[width=\textwidth]{BitcoinExplanationsFrancois_Overview_cropped}
%		\caption{A blockchain is formed of a sequence of blocks containing transactions. The current state of the blockchain (here $(B_0,B_1,B_2)$) is stored by each individual miner.}
%		\label{fig:overview:blockchain}
%		\vspace{23.9pt}
%	\end{minipage}%
%	\hfill
%	\begin{minipage}[b]{0.49\textwidth}
%		\centering
%		\includegraphics[width=\textwidth]{BitcoinExplanationsFrancois_AddingTransaction_cropped}
%		\caption{To add a new transaction to the current blockchain, a miner first verifies the validity of the transaction. It must then solve a costly cryptopuzzle to encapsulate this transaction in a new block (here $B_3$), before disseminating this block to other miners.}
%		\label{fig:adding:a:transaction}
%	\end{minipage}
%\end{figure*}

A blockchain is a \emph{decentralized ledger} composed of blocks containing transactions.
% Each of these data structures are essential to the well functionning of the system.
The transactions, the blocks, and the resulting chain obey a few core rules that ensure the system remains \emph{tamper-proof}.
Great care is required when modifying these rules, as even minor changes might break the blockchain's properties and its security guarantees.%, and lead the whole system to collapse.
% It is essential to understand all of these elements to avoid any safety regression when we attempt to change them.

In the following, we first detail the default workings of the Bitcoin blockchain and its rationale\footnote{Blockchains with closed membership or different consensus protocols are not discussed in this section}.
We then build upon these explanations to introduce and justify the changes we are proposing.

% The three data structures and the policy rules are developed in the following sections.
% A companion data structure, the UTXO set, is also described as it is at the core of the contributions of this paper.\\
%Due to the interlocked composition of Bitcoin, the following technical explanations must be taken as a whole.

\subsection{Overview}

\newcommand{\scaleBitcoinExplanation}{0.39}

\begin{figure}[t]
    \centering
    \includegraphics[scale=\scaleBitcoinExplanation]{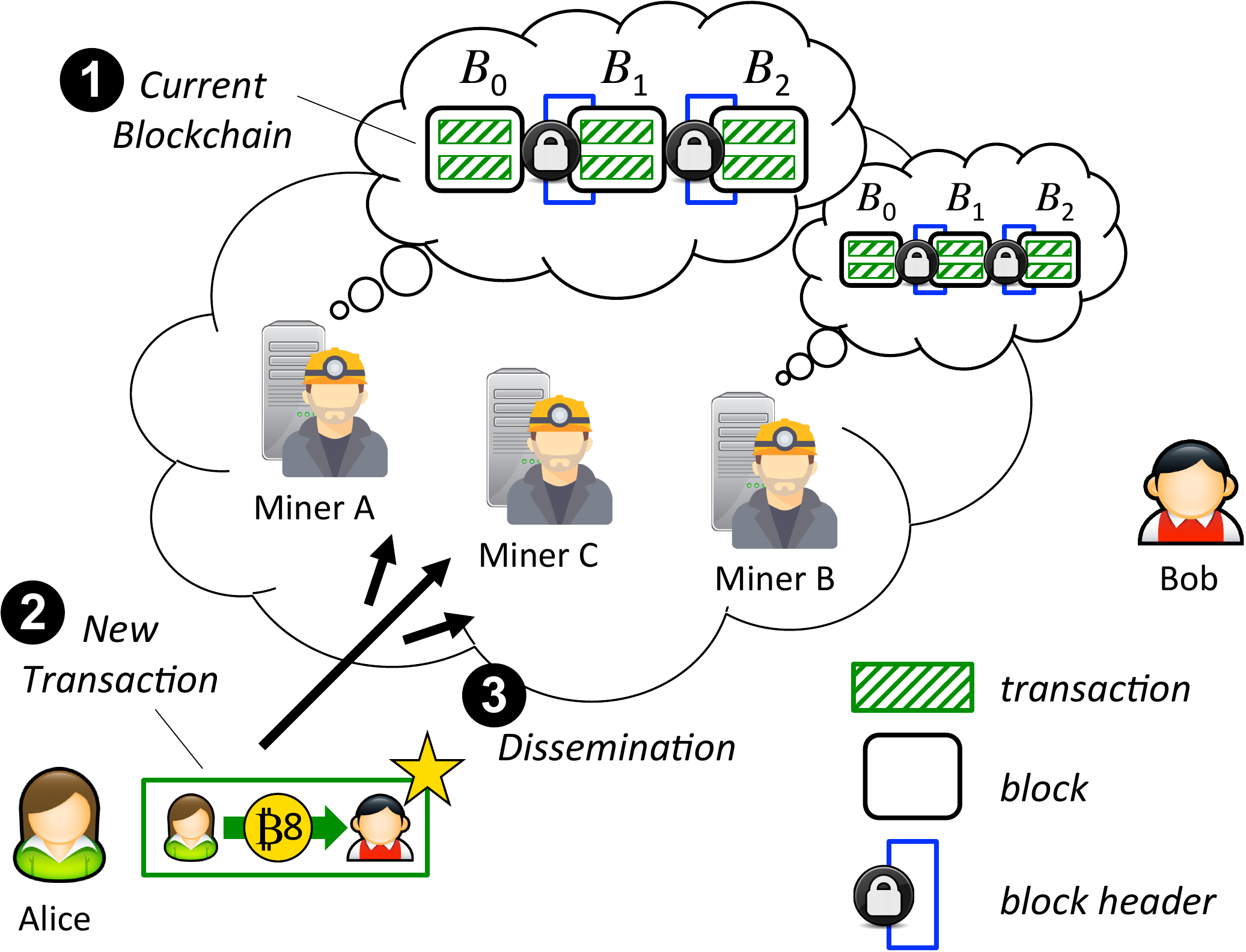}
    \caption{A blockchain is formed of a sequence of blocks containing transactions. The current state of the blockchain (here $(B_0,B_1,B_2)$) is stored by each individual miner.}
    \label{fig:overview:blockchain}
\end{figure}

In a blockchain system such as Bitcoin, the blockchain proper
($(B_k)_{k \in \mathbb{Z}_{\ge 0}}$, label~\circledblack{1} in \figref{fig:overview:blockchain}) is maintained by a peer-to-peer
network of miners. %% It takes the form of a sequence of
%% linked blocks $(B_k)_i$ stored by each miner, with blocks containing
%% the transactions recorded in the blockchain so far. 
Each block $B_k$
links to the previous block $B_{k-1}$
% (shown on the left in the figure)
by including in its header a cryptographic hash that is \emph{(i)}
easy to verify, but \emph{(ii)} particularly costly to create (this second
point is one of the central element of blockchains with open membership, which we will
discuss in detail just below). The leftmost block $B_0$ is known as
the \emph{Genesis Block}: it is the first and oldest block in the
blockchain, and it is the only block with no predecessor.

\subsubsection{Recording a new transaction}

\begin{figure}[t]
    \centering
    \includegraphics[scale=\scaleBitcoinExplanation]{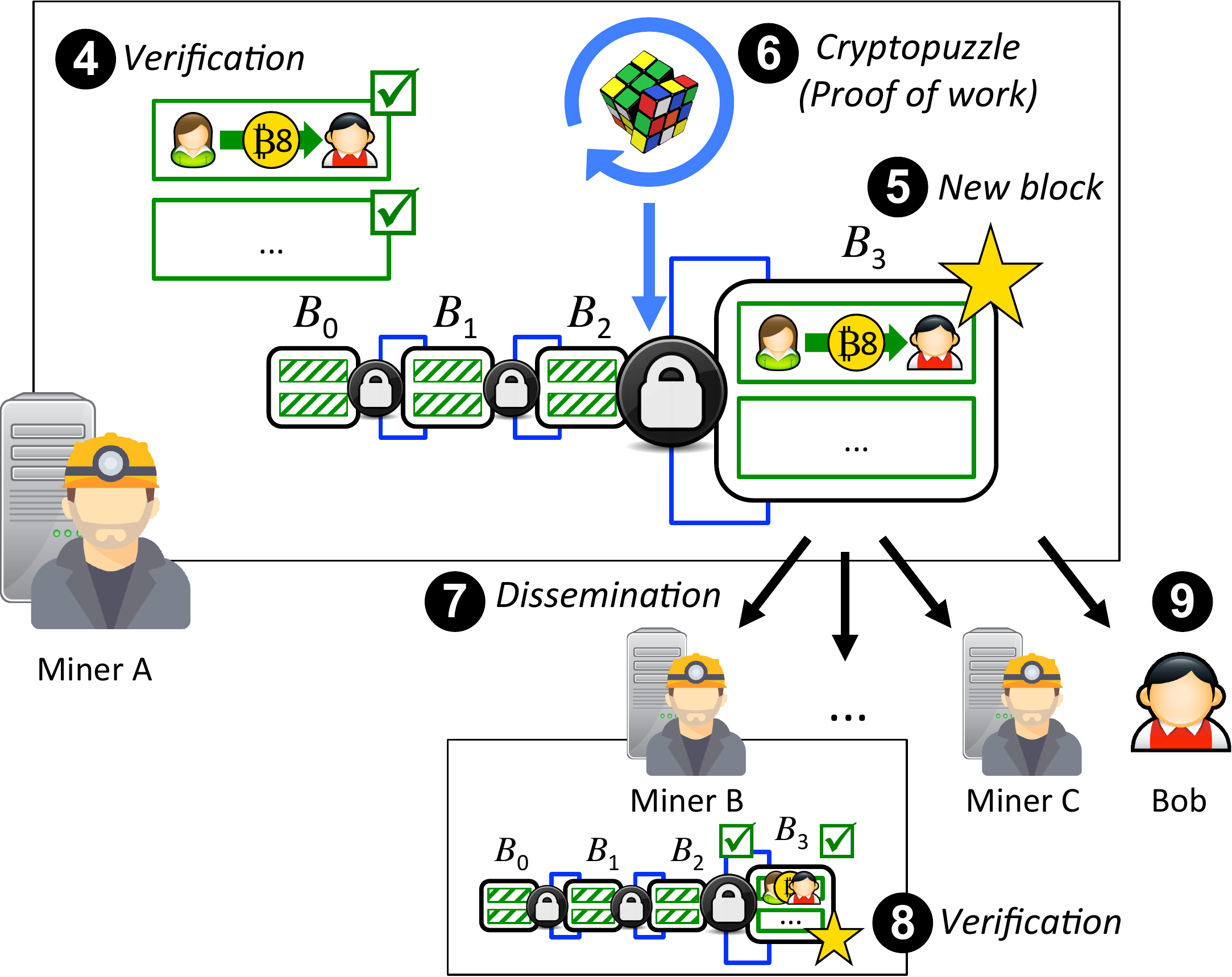}
    \caption{To add a new transaction to the current blockchain, a miner first verifies the validity of the transaction. It must then solve a costly cryptopuzzle to encapsulate this transaction in a new block (here $B_3$), before disseminating this block to other miners.}
    \label{fig:adding:a:transaction}
\end{figure}

To transfer 8 bitcoins from herself to Bob, the user Alice must first create a valid transaction (label~\circledblack{2}) that contains information proving she actually owns the 8 bitcoins (with a cryptographic signature using asymmetric keys), and encode the resulting transaction output with Bob's public key (such that, in turn, only Bob will be able to demonstrate ownership of the transaction's output).

Alice then broadcasts this new transaction to the network of miners
\circledblack{3}, in order for it to be included in the
blockchain. Before adding Alice's transaction into the blockchain,
Miner A first verifies that the transaction is valid
(label~\circledblack{4} in \figref{fig:adding:a:transaction}, details on the transaction verification process will follow in \secref{sec:block}-(\textbf{BV2})).

%% ,
%% i.e. that \emph{(i)} the coins being spent (the transaction's inputs)
%% have been created earlier on in the blockchain; \emph{(ii)} that Alice
%% does indeed own the coins she is spending, and that \emph{(iii)} these
%% coins have not yet been spent (thus preventing Alice from spending her
%% coins twice).

Miner A then includes Alice's transaction together with other
transactions received in parallel into a new block ($B_3$,
\circledblack{5}), and attempts to link it to the current tip of the
blockchain. This linkage operation requires Miner A to solve a
probabilistically difficult cryptopuzzle \circledblack{6} that regulates the frequency at which blocks are created (or \emph{mined}) by the
whole network. (In Bitcoin, this periodicity is set to one block every
10 minutes.) If Miner A succeeds, the new block $B_3$ is now linked to
the existing block chain $(B_0,B_1,B_2)$, and is disseminated to the
other miners \circledblack{7}. When a miner receives a new block (here
Miner B), it checks that the transactions included are valid, along
with the cryptopuzzle \circledblack{8}, before including the new block in its local
copy of the blockchain. The new block ultimately reaches Bob \circledblack{9}, who can check that
the transaction has been properly recorded (and can then, for example,
sell some goods to Alice).

%% Miners are incentivized to produce blocks by receiving some new bitcoins for each new block (which constitutes bitcoin's money creation mechanism), and by receiving fees from users (such as Alice here) whose transactions they have successfully recorded in the chain.

%\subsubsection{Handling divergence and branches in the chain}
\subsubsection{Irrevocability of deep blocks}

%% \begin{figure}[t]
%%     \centering
%%     \includegraphics[scale=\scaleBitcoinExplanation]{BitcoinExplanationsFrancois_BranchedChain_cropped}
%%     \caption{When a branch occurs, only the longest chain (measured in terms of ``work'') is considered valid.}
%%     \label{fig:branch:in:chain}
%% \end{figure}

Because blocks are produced at a limited rate  such that all the miners receive block $B_k$ before they can successfully mine a concurrent block $B_{k'}$, honest miners are highly likely to extend the chain when producing a new block, ensuring a consistent system state with high probability.
The views of individual miners may however diverge in problematic cases, causing
branches to appear. %% For instance in \figref{fig:branch:in:chain},
%% a first blockchain $(B_0,B_1,B_2)$ has been constructed by the system,
%% but two new blocks $B_{3a}$ and $B_{3b}$ have then been produced that
%% link back to $B_2$. The system now contains two competing chains
%% $(B_0,B_1,B_2,B_{3a})$ and $(B_0,B_1,B_2,B_{3b})$. Such branches might
%% appear if two miners succeed in creating a block almost simultaneously
%% (an unlikely albeit not impossible event), or due to a partition in
%% the network, causing one set of miners to become unaware of the work
%% of the rest of the system.
When a branch occurs, miners resolve the divergence by choosing as valid branch the one that was the most difficult to create (details on block difficulty will follow in \secref{sec:block}-(\textbf{BV2})). %the longest branch\footnote{The best branch is in practice the branch with the highest sum of all its block difficulty.} as the valid one.
Blocks that are left out of the chain are said to be \emph{orphan}. %% In \figref{fig:branch:in:chain} because an additional block $B_4$ has been attached to $B_{3b}$, the lower branch is consider the valid blockchain. As a result, the block $B_{3a}$ (and the transactions it contains) is no longer considered to be part of the chain, and is said to be \emph{orphan}.

\begin{figure}[t]
    \centering
    \includegraphics[scale=\scaleBitcoinExplanation]{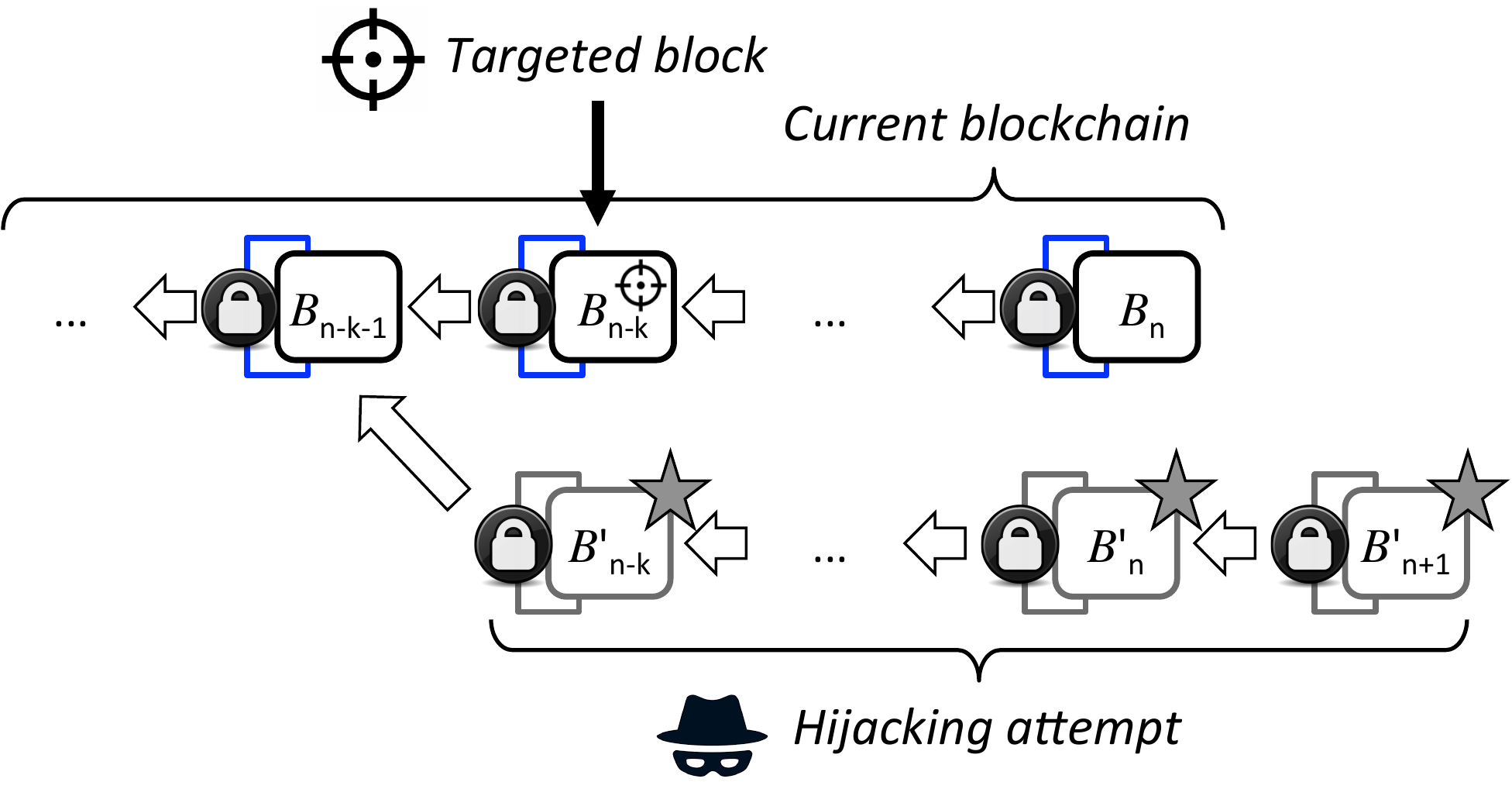}
    \caption{Revoking the content of a block $B_{n-k}$ deep in the chain requires constructing a better alternative subchain, which becomes exponentially harder as the block lies deeper.}
    \label{fig:highjacking:attempt}
\end{figure}

%% Blocks that are close to the tail of the chain %% (such as $B_{3a}$ and
%% %% $B_{3b}$ here)
%% always run the risk of being made orphan by a new
%% branch.
The risk of being made orphan decreases exponentially as a block lies
deeper in a chain, ensuring the \emph{practical irrevocability} of
deep blocks and the transactions they contain. This is illustrated in \figref{fig:highjacking:attempt}: consider an attacker who wishes
to revoke a block $B_{n-k}$ (targeted block), that lies $k$ blocks
away from the chain's tip $B_n$. For this attack to succeed, this
attacker must produce an alternative subchain $(B'_{n-k}, .., B'_{n},
B'_{n+1})$ that is more difficult to create than the current chain. Producing this
subchain is however extremely costly, and takes time which increases the
odds that the legitimate chain grows (with a block $B_{n+1}$,
thus requesting an even more difficult attack subchain) before the attacker
succeeds. When the computing power of the attacker is less than half
of that of the rest of the network, his probability of success drops
exponentially with $k$.  

% Miners are incentivized to 
%\subsection{Data structures}
\subsection{Transactions, Blocks, and UTXO set}

To benefit from the full security of Bitcoin, Bob should verify the
validity of the new block that contains Alice's payment to him
(label~\circledblack{9} in \figref{fig:adding:a:transaction}) in
addition to verifying the validity of Alice's transaction. This is
because Alice could collude with a miner (or launch herself a miner),
and produce an invalid block that she would advertise to Bob.  Bitcoin
relies on a number of built-in validity checks on blocks and
transactions to conduct this verification.  However, whereas
\emph{full nodes} exploit all of these checks, \emph{Simple Payment
  Verification nodes} (SPV nodes) only perform a limited verification.
In the following, we describe the details of these validity checks, we discuss the role of an intermediary set known as the set of \emph{Unspent TransaCTion Outputs} (UTXO set), and the shortcomings of SPV nodes ensued by the limited verification they perform.

%% The depth of the
%% verification performed by Bob depends on whether Bob runs a \emph{full
%%   node} or an \emph{Simple Payment Verification} (SPV) node as we now
%% discuss in more detail. 

%\todo{Probably inject something about need to check as early as possible, as useful, since an alternative could be for Bob just to wait for the block to drop deeper in the chain}
%\somenote{Stopped here FT29Sep17}
% what role

\subsubsection{Checking block validity}
\label{sec:block}

A block is valid if and only if it meets the following two conditions.
\begin{itemize}
\item {\bf (BV1)} Its header respects the blockchain's \emph{Proof-of-Work predicate}.
\item {\bf (BV2)} It only contains valid transactions (which we discuss further below).
\end{itemize}

\begin{figure}[t]
  \centering \includegraphics[scale=\scaleBitcoinExplanation]{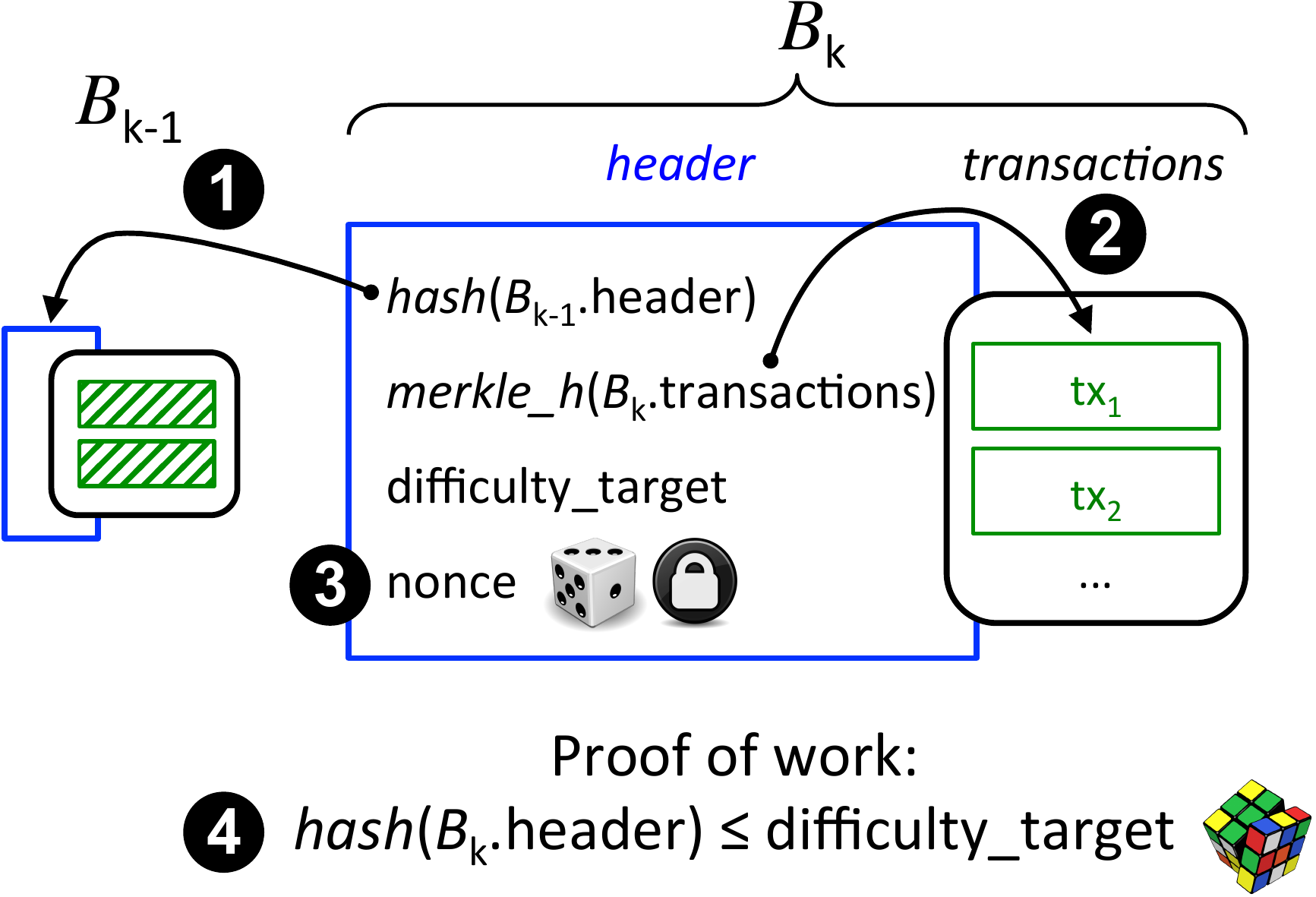}
  \caption{Content of a block header (simplified).}
  \label{fig:block:header}
\end{figure}

\noindent \textbf{BV1}: The Proof-of-Work predicate makes it very difficult for malicious actors to alter the blockchain in an attempt to edit the ledger.
The Proof-of-Work predicate is used as a lock-in mechanism to anchor blocks in the chain.
It is enforced on each block header, whose simplified structure is shown in \figref{fig:block:header}.
%\noindent \textbf{BV1}: The Proof-of-Work predicate makes it very difficult
%to tamper a block's header, whose simplified structure is shown in \figref{fig:block:header}.
The header of each block $B_k$ points both to the header of
the previous block $B_{k-1}$ (using a hash function,
\circledblack{1}), and to the transactions contained in the current
block $B_k$ \circledblack{2}.
To fulfill the Proof-of-Work predicate~\circledblack{4}, a header must contain a \emph{nonce} \circledblack{3} such that the hash of the header is less than a \emph{difficulty target}.
The difficulty target is set so that a new block is created every ten minutes by the miners as a whole, regardless of the computation power (the difficulty target is regularly adjusted to cope with changes in their computation power).
Finding a nonce respecting the difficulty target is computationally very expensive, as every miner competes to create blocks.
This computing cost prevents attackers from easily tampering the chain as they have to recompute fresh nonces for the blocks they wish to replace.%, which requires time and computing power.
% : any modification of the the block's header will break the predicate with high probability, while any modification of the header's transactions will break the Merkle tree hash. In both cases, the block becomes invalid.

\begin{figure}[t]
    \centering
    \includegraphics[width=0.5\textwidth]{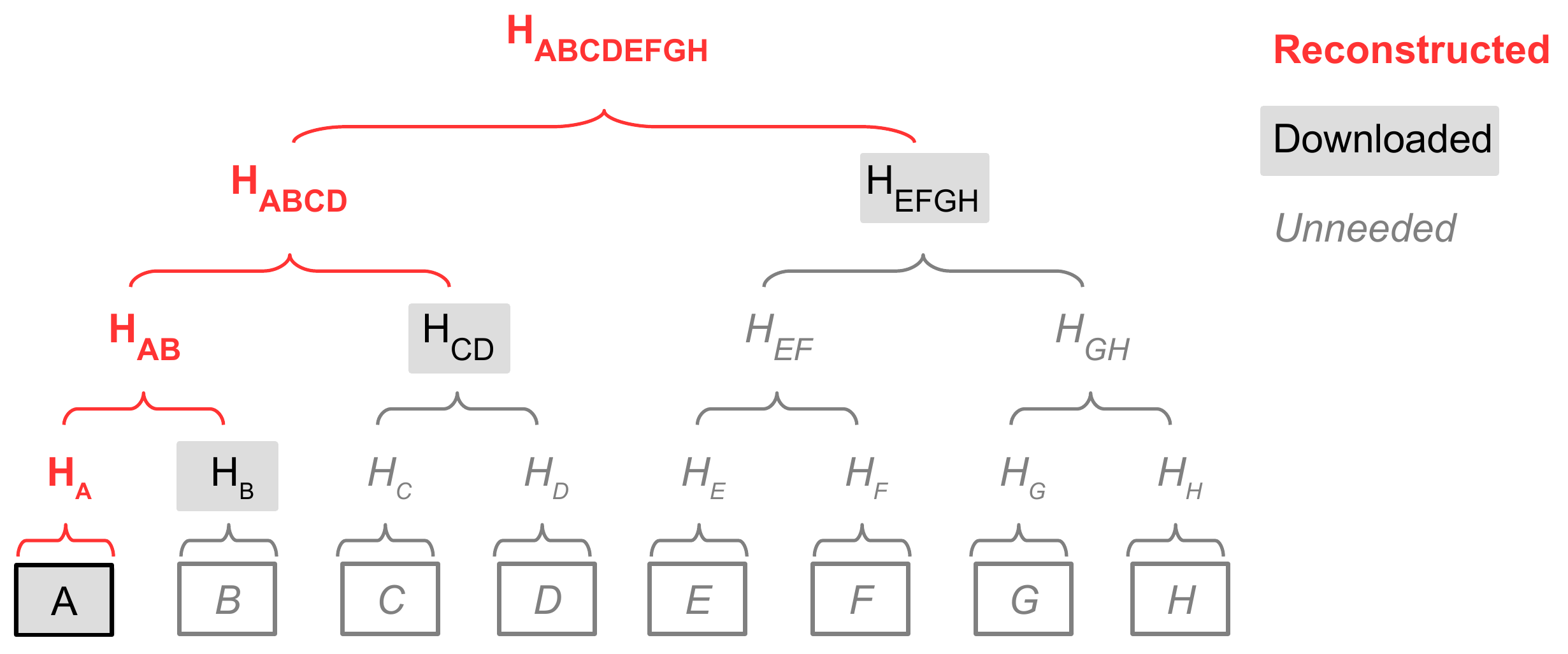}
    \caption{Example of a Merkle tree root reconstruction that only needs $log(n)$ hashes.}
    
    \label{fig:merkle_tree}
\end{figure}

To establish a secure and verifiable link between the header of $B_k$ and the
corresponding block $B_k$, the pointer to $B_k$'s transactions
\circledblack{2} consists of the root of a Merkle tree.
A Merkle tree is a hierarchical hashing mechanism for sets that enables a verifier to efficiently test whether an item (here a transaction) belongs to the set by reconstructing the root of the Merkle tree.
Each leaf node in a Merkle tree consists of the hash of an item, while each internal node (including the root) consists of the hash of its children.
This makes it possible to reconstruct the root, and thus verify set membership, using only a logarithmic number of intermediate hashes.
In \figref{fig:merkle_tree} for example, a node can verify the presence of transaction $A$ in the set by \emph{(i)} downloading the root from a secured communication channel (e.g., the blockchain), and \emph{(ii)} downloading $A$ and the three intermediate hashes shown in red: $H_B$, $H_{CD}$ and $H_{EFGH}$, and reconstructing the root from the downloaded hashes.
The reconstructed root should match the downloaded one.

%  This makes
% it possible to verify the existence of an item $x$ by a partial Merkle
% tree is constructed from the hash of $x$ and from the
% $\log(|dataset|)$ hashes that complement the hash of $x$ in order to
% compute the root of the tree.  If the root of the partial tree is
% equal to the root of the full tree, then $x$ is proven to be in the
% dataset.  In \figref{fig:merkle_tree} for instance, $A$ and 3 hashes
% are sufficient to reconstruct the root of the tree for this dataset of
% 8 items.

 % uses a Merkle tree hash,  % (we provide more details below).

%% This proof-of-work mechanism is also used to throttle the rate at which blocks are being produced by the overall network (1 block every 10 minutes for Bitcoin), making the system almost synchronous. To account for changes in the network the difficulty target is adjusted on a regular basis (typically decreased, thus making the difficulty harder), and included in the header to allow for self-contained verification.

\begin{figure}[t]
    \centering    \includegraphics[scale=\scaleBitcoinExplanation]{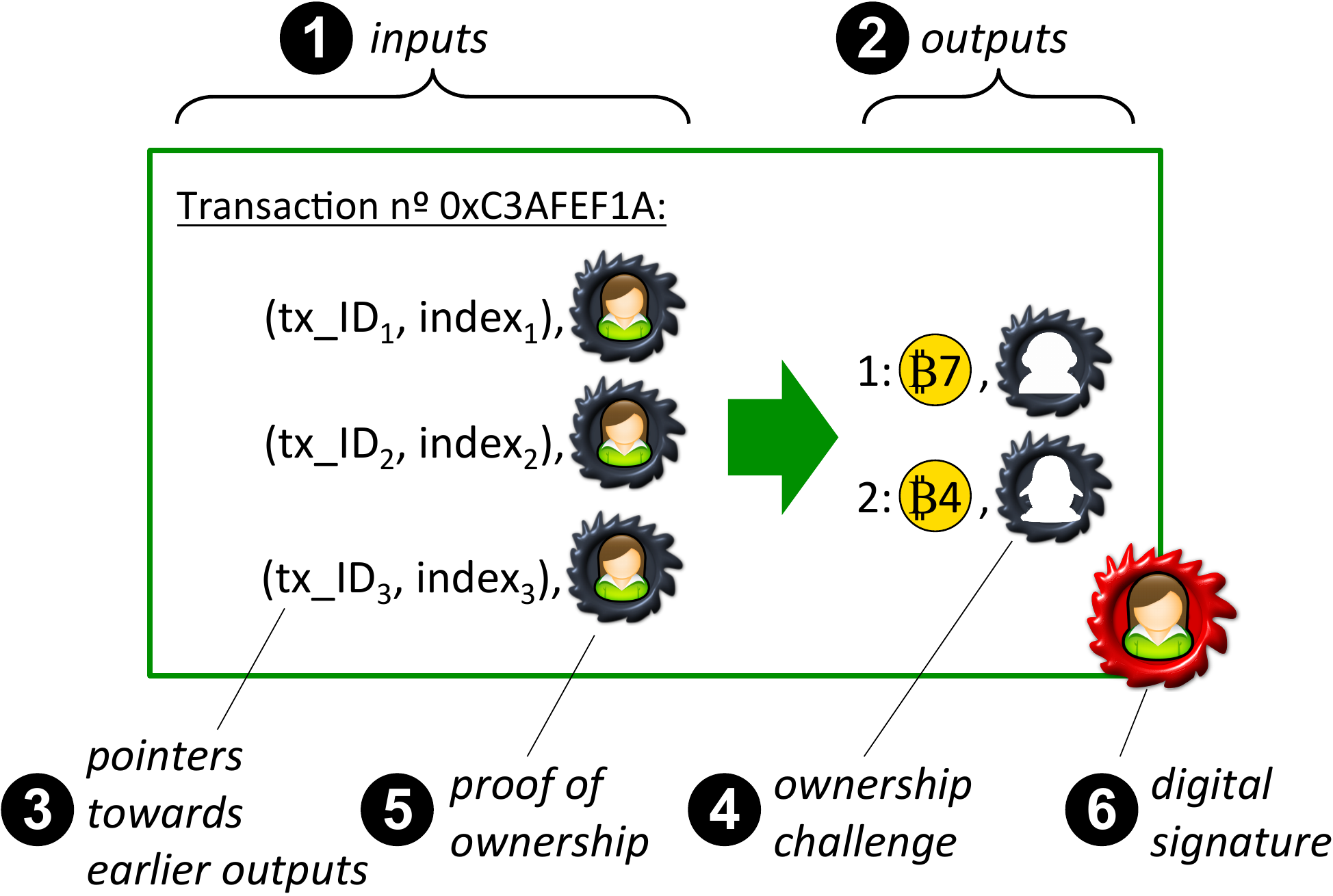}
    \caption{Structure of a transaction.}
    \label{fig:transaction}
\end{figure}

\noindent \textbf{BV2}: 
\label{sec:transaction}
In addition to the Proof-of-Work predicate ({\bf BV1}), all the
transactions included in a block must also be valid for the overall
block to be valid.
%% The validity of a transaction is a necessary precondition for
%% recording the transaction in the blockchain: honest miners will only
%% include valid transactions in a block, and only accept blocks
%% containing valid transactions. % A transaction being valid is however
%% %% not sufficient to ensure it is recorded: it must in addition be
%% %% included in a valid block.  
\figref{fig:transaction} shows the validity mechanisms included in a typical Bitcoin transaction. In this example, Alice uses 3 coins she owns (the transaction's inputs
\circledblack{1}) to pay 7 Bitcoins to Bob, and 4 to Tux (the transaction's outputs \circledblack{2}).

Only coins created in
earlier transactions may be spent: each of Alice's inputs therefore
points back to the output of an earlier transaction
\circledblack{3}. To ensure that only the recipients (Bob and Tux)
are able to spend the output, each new coin contains an ownership
challenge (a hashed public key),
that must be solved to spend this coin
\circledblack{4}.

Alice's transaction is only valid if the following three conditions are met:
\begin{itemize}
\item {\bf (TV1)} The inputs do exist, and Alice owns them. She can prove her ownership of the inputs by providing a public key matching their ownership
  challenges \circledblack{5}, and by signing the new transaction with
  the corresponding private key\footnote{Bitcoin uses a scripting
    language to encode challenges and proofs of ownership, enabling
    for more complex schemes, but for ease of exposition we limit
    ourselves to the typical case.}  \circledblack{6} ;
\item {\bf (TV2)} No money is created in the transaction. In effect, the total value of the transaction's inputs must be greater than or equal to that of its outputs:
  $$\sum_{\substack{\mathit{in}\in \mathsf{inputs}(t)}}{\mathsf{value}(\mathit{in})} \geq
    \sum_{\substack{\mathit{out}\in \mathsf{outputs}(t)}}{\mathsf{value}(\mathit{out})}.$$

The difference $\sum{\mathsf{value}(\mathit{in})} - \sum{\mathsf{value}(\mathit{out})}$ is given as a fee to the miner of the block containing the transaction;
\item {\bf (TV3)} The transaction's inputs $(\mathsf{tx\_ID}_i,\mathsf{index}_j)$ have not been spent yet (i.e., they do not appear as inputs of any earlier transaction, an attack known as a \emph{double spend}).
\end{itemize}

\subsubsection{The set of Unspent TransaCTion Outputs (UTXO set)}
\label{sec:utxo}
While the validity of a block's header ({\bf BV1}) only requires
access to the current block $B_k$, and to the header of its
predecessor $B_{k-1}$, verifying transactions ({\bf BV2}) requires a
lot more information. Verifying the ownership challenges of input
coins ({\bf TV1}), and their amount ({\bf TV2}) requires access to the
transactions recorded in earlier blocks. Worse, verifying that inputs
coins have not yet been spent ({\bf TV3}) potentially requires
parsing and verifying the entire blockchain.

To avoid performing such a costly operation for each new block, nodes that verify transactions maintain an intermediary set known as the set of \emph{Unspent TransaCTion Outputs} (UTXO set).
The UTXO set contains all the coins that have been created in the chain but not spent in later transactions, it thus contains all the spendable coins.

A node verifying transactions can prevent double spends ({\bf TV3}) by simply ensuring that all the inputs of a transaction appears in its UTXO set.
The UTXO set evolves as new correct blocks are added to the chain: transaction outputs are removed from the set when they are spent, and outputs of new transaction are added to the set.

%% As you can see from \secref{sec:transaction}, coins are moved from one transaction to another, unlike classical transfers where money is moved from one account to another.
%% Thus, all the available coins are stored within a subset of the transaction outputs in the blockchain.
%% Personal accounts, called wallets, merely contain pointers to the transaction outputs of its owner.

%% The set of \emph{Unspent TransaCTion Outputs} (UTXO) represents all the available coins that are ready to be used as inputs in future transactions.

%The UTXO set is different from one block to another, it can be seen as a snapshot of the coins available to be spent

\subsection{The limitations of SPV nodes}

In spite of its benefits, constructing a local UTXO set is costly: in order to obtain the set, a node must first download the entire chain (120 GiB as of August 2017, see \figref{fig:sizes} and validate it (a lengthy process that can take hours on high-end machines), even if only the latest block is relevant to its interest.

Because of this cost, Bitcoin supports several levels of verification. \emph{Miners} and users running \emph{full nodes} construct the UTXO set and check both block headers ({\bf BV1}) and transactions ({\bf TV1,2,3} and as a result {\bf BV2}). By performing all the possible checks, full nodes benefit from the maximum security that the Bitcoin system has to offer.

By contrast, \emph{Simple Payment Verification} nodes (SPV nodes) do not construct the UTXO set. Instead, they only download the chain's block headers (rather than full blocks), and verify that these headers are valid ({\bf BV1}). With the headers only, SPV nodes are able to verify the well-formedness of the blockchain including the crucial Proof-of-Work predicate that seals the links of the chain. %, which are the strongest properties that Bitcoin offers against tampering attacks.

% Low resource devices in Bitcoin are typically SPV nodes and as thus only synchronize and perform verification on the blocks headers.

However, because this verification is only partial, SPV nodes are unable to detect if a new block contains an invalid transaction. This scenario, however probabilistically difficult to accomplish for an attacker, is a vulnerability of SPV nodes. To circumvent this vulnerability, SPV nodes typically wait until miners have created subsequent blocks extending the chain containing a block of interest, which implies that these miners have performed a full verification on it and consider this block as valid.

The need for SPV nodes to wait makes it particularly problematic to use Bitcoin on limited devices (i.e. mobile phones) for everyday transactions. SPV nodes are not even able to check whether the inputs used in a transaction do exist. It also limits the ability of SPV nodes to detect faulty transactions as early as possible, which is an important usability feature of modern payment systems.

In this work, we propose to overcome the inherent limitations of SPV nodes with \emph{Dietcoin}. Dietcoin enables nodes with limited resources (\emph{diet nodes}) to benefit from a security level that is close to that of full nodes, at a fraction of the cost required to run full security checks. 

% limit ability to detect early an important feature of financial transactions

%% However, with only partial knowledge of blocks, SPV nodes can only perform partial verification on blocks. For instance, if a block were to contain an invalid transaction (e. g., that grants its spender a large amount of coins) but a correct hash, an SPV node would classify the block as valid whereas a full node would rightfully classify it as invalid. This scenario, however probabilistically difficult to accomplish for an attacker, is a vulnerability for SPV nodes. To circumvent this vulnerability, SPV nodes typically wait until miners have created other blocks on top of a block of interest, which imply that these miners have performed full verification on it and consider this block as valid.

%% SPV nodes that are not willing to wait to receive additional blocks are vulnerable to these forged blocks.

%%% Local Variables:  %%% mode: latex %%% TeX-master: t %%% End: 

%%  LocalWords:  cryptographic blockchain irrevocability verifier

%% file: dietcoin_system.tex
% -*- tex-main-file: "RR-9162" compile-command: "make pdf" ispell-dictionary: "american" -*
\newcommand{\utxo}{\ensuremath{\textsc{UTXO}}}
\section{The Dietcoin system}
\label{sec:dietcoin}
 To address the vulnerabilities of
SPV nodes and to improve the confidence mobile users can have in
recent transactions, we propose Dietcoin, an extension to Bitcoin-like
blockchains.
Although our proposal can be applied to most existing
Proof-of-Work blockchains using the UTXO model for coins, we describe Dietcoin in the context
of the Bitcoin system as presented in \secref{sec:bitcoin}.

The core of Dietcoin consists of a novel class of nodes, called \emph{diet
  nodes}, which provide low-power devices with the ability to perform
\emph{full block verification} with minimal bandwidth and storage requirements.
%Diet nodes can increase their trust in the
%transactions they are interested in by downloading and verifying the corresponding
%blocks with minimal bandwidth and storage requirements.
Instead of having to download and process the entire
blockchain to build their own copy of the UTXO set, diet nodes query the UTXO set of full nodes and
use it to verify the legitimacy of the transactions
 they are interested in (as described in \secref{sec:block}-({\bf BV2})) and the correctness of the blocks that contain them.
%use it to perform the transaction verification (as described in \secref{sec:block}-({\bf BV2})) on
This gives diet nodes
security properties that sit in between those of full nodes, and those
of Bitcoin's SPV nodes.

Consider a user wishing to verify a transaction for the sale of some
goods. The user's diet node will initially proceed like a standard SPV
node. It will contact a full node to obtain the header of the block
that supposedly contains its transaction as well as the corresponding
branch of the transaction Merkle tree, to verify that the transaction
indeed is included in the block. But while an SPV node would stop at this
inclusion check, the diet node continues by verifying both the inclusion and the correctness of all the transactions in the block.
To make this possible we introduce the possibility for diet
nodes to access the state of the UTXO set of full nodes corresponding to the instant
right before the block they want to verify.

% verify 1 block, n block, dl utxo, verify old tx by DL 1 shard
Since downloading the
entire UTXO set would result in prohibitive bandwidth overhead, as shown in \figref{fig:sizes},
Dietcoin-enabled full nodes split their UTXO set into small shards,
enabling diet nodes to download only the shards that are
relevant to the transactions in the block.  

To prevent diet nodes from trusting maliciously forged shards, the shard hashes are used as leaves of a Merkle tree, which root is stored in each block.
Having the Merkle root stored in blocks enables the UTXO shards to benefit from the same Proof-of-Work protection as transactions.

In addition to the full verification done on the block of their interest $B_k$, diet nodes can increase their trust in $B_k$ by fully verifying its previous blocks.
By doing so, diet nodes are ensured that none of the $l$ verified blocks contain illegal transactions nor erroneous UTXO Merkle root.
To make a diet node trust a forged transaction in block $B_k$, an attacker has to counterfeit the subchain of $l+1$ blocks $(B_{k-l}, ..., B_k)$.
Thanks to the Proof-of-Work protection, the cost of this attack increases exponentially as $l$ increases linearly.

% Downloading UTxO shards allows a diet node to verify legitimacy of a
% block's transactions, based on the assumption that the shards are
% correct. But Dietcoin also enables diet nodes to iterate the
% verification to make sure that the downloaded UTxO shards are also
% sound. In the following, we detail the operation of Dietcoin by first
% describing how full nodes and miners can provide diet nodes with
% verifiable UTxO shards. We then discuss how miners link blocks to
% their UTxO state. Finally, we detail the operation of diet nodes when
% verifying transactions.

In the following, we detail the operation of Dietcoin by first
describing how full nodes can provide diet nodes with
verifiable UTXO shards. Secondly we discuss how miners link blocks to
the state of their UTXO set.
We then explain how diet nodes can extend the verification process from one block to a subchain of any length.
Finally, we detail the operation of diet nodes when
verifying transactions.

% For the verification process to be sound, it is important for diet
% nodes to verify not only the transactions in block, but also the
% legitimacy of the UTXO shards they use in the verification process. To
% this end, dietcoin-enabled full nodes organize the UTXO shards into a
% UTXO Merkle tree, and store the root hash of this tree inside the
% blocks.

% Querying UTXOs to a remote nodes is a simple task, but how do Diet nodes ensure the legitimacy of the downloaded UTXOs?
% The core of the technical contributions is a system design that enables Diet nodes to trust the UTXOs they download and reject the corrupted UTXOs.
\subsection{Sharding the UTXO set}
\label{sec:sharding-utxo}
To enable the operation of diet nodes, Dietcoin-enabled full nodes
need \emph{(i)} to provide diet nodes with shards of the
UTXO set, while \emph{(ii)} enabling them to verify that these shards are
authentic. 

To satisfy \emph{(i)}, Dietcoin-enabled full nodes store the UTXO set resulting
from the application of the transactions in each block in the form of
shards with a predefined maximum  size of 1\,KiB (on average, across all shards).
The use of shards enables
diet nodes to download only the relevant parts of the UTXO set and also
limits the storage requirements at full nodes, which only need to
store the modified shards for each block to let diet nodes query older versions of the shards. Similarly, the limit of
1\,KiB for the size of each shard limits the bandwidth employed by diet
nodes in the verification process.

To satisfy \emph{(ii)}, full nodes also maintain a Merkle tree that indexes
all the shards of the UTXO set. Using shards also proves advantageous with
respect to this Merkle tree. If nodes were to index UTXO entries
directly, the continuous changes in the UTXO set would cause the
Merkle tree to become quickly unbalanced, leading to performance
problems or requiring a potentially costly self-balancing tree.  The
use of shards, combined with the right sharding strategy, gives the UTXO Merkle tree a
relatively constant structure, enabling shards to be updated in place
most of the time.
Moreover, it makes it possible
to predict the size of the UTXO Merkle tree and its incurred overhead,
which enables us to better control and balance the storage overhead
for full nodes and the bandwidth requirements of diet nodes.

A number of sharding strategies satisfy the requirement of a fixed
number of shards. In this work, we use the simplest approach
consisting of indexing UTXO entries by their first $k$ bits.  This
strategy resembles a random approach since the first bits of an UTXO are the transaction hash it references, which value is expected to be random due to the uniformity property of
the SHA-256 hash function. This strategy comes with the added
advantages of obtaining shards of homogeneous size and a full binary
Merkle tree with $2^k$ leaves. 

Keeping in mind the size cap of 1\,KiB per shard, $k$ can be adapted locally by each node to cope with the growth of the UTXO set.
When the average shard size breaches the cap of 1\,KiB, $k$ is incremented by one resulting in \emph{(i)} halving the average shard size, and \emph{(ii)} adding one layer to the Merkle tree, doubling its storage footprint in the process.

\subsection{Linking blocks with the UTXO set}
\label{sec:verify-trans}
\begin{figure}
	\centering
	\includegraphics[width=0.5\textwidth]{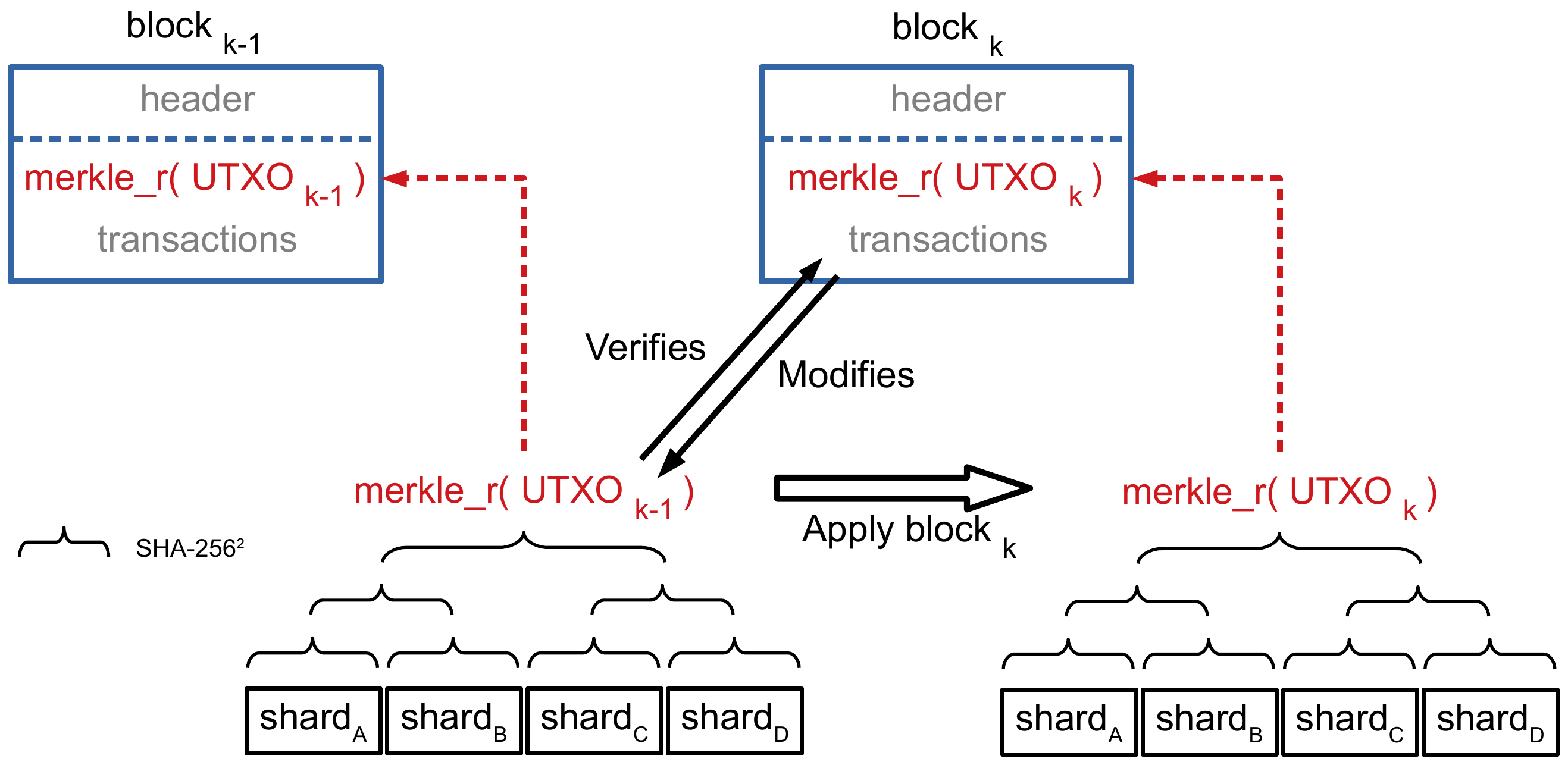}
	\caption{The UTXO set is updated every time a block is validated. For a counterfeited block to be validated by diet nodes, a malicious node has to forge at least two consecutive blocks: the first block $B_{k-1}$ containing a fake Merkle root of $\utxo_{k-1}$, and the second block $B_k$ spending fake coins validated by the fake $\utxo_{k-1}$.}
% present in $B_{k-1}$.}
	\label{fig:utxo_snapshot}
\end{figure}

With reference to \figref{fig:utxo_snapshot}, consider a Dietcoin-enabled miner that
is mining block $B_k$, and let $\utxo_k$ be the state of the UTXO set
after applying all the transactions in $B_k$. The miner stores the
root of the UTXO Merkle tree associated with $\utxo_k$ as an
unspendable output in the first transaction of block $B_k$
before trying to solve the Proof-of-Work as shown
%in Figures~\ref{fig:block_changes} and~\ref{fig:utxo_snapshot}.
in \figref{fig:utxo_snapshot}.
Storing the root of the UTXO Merkle tree in the first transaction of the
block does not require any modification to the structure of Bitcoin's
blocks. Thus it is possible for Dietcoin-enabled full nodes and
miners to co-exist with their legacy Bitcoin counterparts.

Dietcoin-enabled full nodes verify the value of the UTXO Merkle root
against their own local copy when they verify a block, while legacy
Bitcoin nodes simply ignore it.

%\begin{figure}[t]
%    \centering
%    \includegraphics[width=0.5\textwidth]{block_changes}
%    \caption{When a block is created, the root of the UTXO Merkle tree is integrated in the block for diet nodes to later verify the validity of the UTXO shards they download.} 
%    \label{fig:block_changes}
%\end{figure}

The UTXO Merkle tree provides a computationally efficient way for diet
nodes to verify whether the shards they download during the
verification process are legitimate and correspond to the current
state of the ledger.  Still referring to \figref{fig:utxo_snapshot}, let us
consider a diet node $d$ that wishes to verify a transaction in block
$B_{k}$. Node $d$ needs to obtain: block $B_k$, the UTXO Merkle root in block $B_{k-1}$,
the shards of the UTXO set between the two blocks, and the elements of the
associated Merkle tree that are required to verify their
legitimacy. It can then verify the shards using the root stored in block
$B_{k-1}$'s first transaction, and use them to verify the correctness of the
transactions in $B_k$.

We observe that storing the Merkle-root referring to the state
after the block into the block itself forces diet nodes to download
two blocks to verify transactions. This makes it inherently harder for
an attacker to provide a diet node with a fake block $B_{k}$ because
it would need to forge not only block $B_{k}$ but also block
$B_{k-1}$.

% Moreover, diet nodes can decide to extend their confidence in
% transactions by iterating the verification process towards previous
% blocks for $\lv{maxBlocks}$ times . In the case of \figref{fig:utxo_snapshot},
% assuming $\lv{maxBlocks}=2$, diet node $d$ will download block
% $B_{k-1}$ to verify the transactions and the UTXO Merkle root in $B_{k}$
% further increasing its confidence in $B_{k+1}$.

\subsection{Extended verification}
\label{sec:extended_verification}
Diet nodes have the ability to extend their confidence in a block by iterating the verification process towards its previous blocks.
By doing so, diet nodes ensure the correctness of the UTXO Merkle root present in block $B_{k-1}$ used to verify the correctness of block $B_k$.
% that each block $B_k$ is verified by using a correct UTXO Merkle root present in $B_{k-l}$.
The extended verification can be performed on a subchain of any length $l$ provided that the verifying diet node can query UTXO shards of any age. 
A diet node fully verifying the subchain $(B_{k-l+1}, ..., B_k)$ can only be tricked into trusting a malicious transaction in block $B_k$ if the attacker manages to counterfeit the $l+1$ successive blocks starting from $B_{k-l}$, which becomes exponentially more costly as $l$ increases linearly.
% an attacker has to counterfeit at least $l+1$ blocks to trick an $l$-diet node into trusting a transaction in block $B_k$.

The block $B_{k-l}$ contains the UTXO Merkle root that serves as a basis for the verification of the consequent blocks.
Since the block $B_{k-l}$ is not verified, it is thus trusted by the diet node.
A comparison can be made with full nodes where the first block of the chain, the genesis block, is hard-coded and is therefore trusted.
By shifting the trust from the genesis block to the block $B_{k-l}$ for diet nodes, Dietcoin effectively shortcuts the verification process.

Picking the value of $l$ exhibits a trade-off between security and verification costs.
On one hand choosing a great $l$ draws the behavior of the diet node closer to that of a full node, while on the other hand choosing a small $l$ draws it closer to that of an SPV node.
The user can make her decision based on which block depth $l$ is great enough in her opinion such that all blocks prior to $B_{k-l}$ are unlikely to be counterfeited.
For instance, a diet node user can choose $l$ such that the trusted block has a block depth of 6 or greater since it is the de facto standard in Bitcoin to consider blocks of depth 6 or greater as secured\footnote{\url{https://bitcoin.org/en/developer-guide#verifying-payment}}.

In the case depicted in \figref{fig:utxo_snapshot}, assuming $l = 2$, diet node $d$ downloads block $B_{k-1}$ to verify the transactions and the UTXO Merkle root in $B_{k-1}$ to further increase its confidence in $B_{k}$.
To verify $B_{k-1}$, the diet node uses the UTXO Merkle root in $B_{k-2}$.

\subsection{Detailed Operation}
\label{sec:diet_nodes}
Equipped with the knowledge of Dietcoin's basic mechanisms, we can now
detail the verification process carried out by diet
nodes. \algoref{algo:spv_compact} depicts the actions taken by a diet
node when its user starts the application using Dietcoin and compares
it with those taken by legacy SPV clients.  Black dotted
lines [$\bullet$] are specific to diet nodes, while hollow dotted lines [$\circ$]
are common to both diet and SPV nodes.

The algorithm begins when
the application using Dietcoin starts and updates its view of the
blockchain. In this first part of the algorithm, a diet node behaves
exactly in the same manner as an SPV node. First, it issues a query
containing the latest known block hash and an obfuscated
representation of its own public keys in the form of a bloom filter
(lines 4-5).  A full node responds to this query by sending a list of
all the block headers that are still unknown to the SPV/diet node,
together with the transactions matching the SPV/diet node's bloom
filter, and the information from the transaction Merkle tree that is
needed to confirm their presence in their blocks.  Using this
information, the SPV/diet node verifies the received headers
(including Proof-of-Work verification) and updates its view of the
blockchain (line 6).%, and verifies that the received transactions are
%in the blocks (lines 6-9).

Since the response from the full node might contain false positives due to the use of a bloom filter for public key obfuscation, the next verification step consists in ensuring that the received transactions match one of the user's public keys (lines 8-9).
Once false positives are discarded, the SPV/diet node verifies that the received transactions are in the blocks (line 10-12).

At this point, a standard SPV node simply returns the received
transactions to the application (line 14). A diet node, on the other
hand, continues the verification process. To this end, the diet node
%needs, for each of the latest $\lv{maxBlocks}$, the corresponding
%block content, the state before the block of UTXO shards associated
%with the block's transactions' inputs and outputs, and the UTXO Merkle
%tree nodes required to rebuild the root hash from these shards. In
%addition, it needs the root hash for the UTXO before the block at
%depth $\lv{maxBlocks}$, that is the one contained in the coinbase of
%the block at depth $\lv{maxBlocks} + 1$. The node checks if this
%information is locally available, and downloads the missing pieces by
%contacting a distinct full node for each block.
first computes which blocks to fully verify to ensure that \emph{(i)} no block is fully verified twice (line 17), \emph{(ii)} old blocks considered by the user as secured enough are not fully verified (parameter $\gv{maxDepth}$, line 17) and \emph{(iii)} only a subchain of limited length $l$ is fully verified (parameter $\gv{maxLength}$, line 19).
If no block is selected for full verification, the diet node falls back to SPV mode (lines 20-21).

To bootstrap the full verification process, the diet node must first download the UTXO Merkle root present in the block prior to the first block to verify (line 23).
For each of the selected blocks $B_k$, the diet node
downloads from Dietcoin-enabled full nodes \emph{(i)} the
block $B_k$ itself (line 25), \emph{(ii)} the state, before $B_{k}$, of the UTXO shards associated with both the block's transactions' inputs and outputs (line 26), and \emph{(iii)} the partial UTXO Merkle tree required to prove the integrity of the downloaded shards (line 26).

%the state before the block of UTXO shards associated with the block's transactions' inputs and outputs, and the UTXO Merkle tree nodes required to rebuild the root hash from these shards.
%In
%addition, it needs the root hash for the UTXO before the block at
%depth $\lv{maxBlocks}$, that is the one contained in the coinbase of
%the block at depth $\lv{maxBlocks} + 1$. The node checks if this
%information is locally available, and downloads the missing pieces by
%contacting a distinct full node for each block.

Once it has all the data, the diet node proceeds with the verification
process. For each block $B_k$, it first verifies that the downloaded UTXO
shards match the UTXO Merkle root from the previous
block $B_{k-1}$ (lines 27-29). Then it verifies that the transactions in each block
use only available inputs from these shards (lines 32-33), and computes the new
state of these shards based on the transactions in $B_k$ (lines 34, 36). Finally it verifies
that the updated shards lead to the UTXO Merkle root contained in $B_k$ (lines 37-39).
Once the verification process has terminated, the diet node returns
the transactions associated with the local user to the application (line 14).

%%% Local Variables: 
%%% mode: latex
%%% TeX-master: t
%%% End: 

%% file: dietcoin_related.tex
% -*- tex-main-file: "RR-9162" compile-command: "make pdf" ispell-dictionary: "american" -*

\section{Related work}
\label{sec:related_work}

Making the UTXO set available for queries between nodes has been discussed several times in the Bitcoin community over the past few years.
Bryan Bishop published a comprehensive list of such proposals~\cite{bryan_[bitcoin-dev]_2017} that share some of the following goals: \emph{(i)} enabling faster node bootstrap, \emph{(ii)} strengthening the security guarantees of lightweight nodes, and \emph{(iii)} scaling the UTXO set to reduce its storage cost.
The primary goal of Dietcoin is to strengthen the security of lightweight nodes.
Dietcoin's strongest feature is the ability for diet nodes to efficiently perform subchain verification, as described in~\secref{sec:extended_verification}, which offers stronger security guarantees than the referenced proposals made by the community.
Moreover, even though we focus in this report on the security of lightweight nodes, it is also possible to bootstrap full nodes faster with Dietcoin.% which can verify the validity of the UTXO set they downloaded by perfo.

Sharing similar goals with Dietcoin, Andrew Miller~\cite{miller_storing_2012} suggests to store in blocks the root of a self-balancing Merkle tree built on top of the UTXO set.
In such a system, lightweight nodes only download the UTXOs they need, which results in a lower bandwidth consumption than with shards as we propose it, but at the cost of a greater storage overhead since the stored Merkle tree is larger.
Moreover, Dietcoin combines a full, and thus always balanced, Merkle tree built on top of $2^k$ shards that can each be updated in place as blocks are appended to the chain.
This combination of tree stability and updatable shards enables efficient subchain verification, as described in~\secref{sec:extended_verification}, and adapting this feature to a system using a self-balancing Merkle tree does not seem trivial.

Vault~\cite{cryptoeprint:2018:269} also proposes to use Merkle trees to securely record the state of the distributed ledger in recent blocks, and shards this state across nodes, to reduce storage costs. Contrarily to Dietcoin, however, Vault targets balance-based schemes, such as introduced by Ethereum, in blockchains relying on Proof-of-Stake consensus. Vault further stores individual accounts in the Merkle trees, rather than UTXO shards as we do. This represents a different and to some extent orthogonal trade-off to that of Dietcoin, in that Vault chooses to increase the size of Merkle tree witnesses that must be included in transactions, but removes the need for lightweight nodes to download UTXO shards.

Whereas we focus on sharding the resulting state of the blockchain, other systems propose to shard the verification process.
Both Elastico~\cite{luu_secure_2016} and OmniLedger~\cite{kokoris-kogias_omniledger:_2018} proposes a permissionless distributed ledger using multiple classical PBFT consensus protocols each executing within a subset (shard) of nodes.

Elastico limits the number of shards a malicious nodes may join (under different identifies) by tying the shard of a node to the result of a Proof-of-Work puzzle.
OmniLedger~\cite{kokoris-kogias_omniledger:_2018} extends the ideas proposed by Elastico~\cite{luu_secure_2016} to increase the size of the shards (and thus reduce the probability of failures), and allow for cross-shards transactions thanks to a Byzantine shard-atomic commit protocol called \emph{Atomix}. Both Elastico and OmniLedger use sharding to increase the transaction processing power of a distributed ledger, rather than to improve access and verification of the UTXO set, as we do.

Chainiac~\cite{nikitin_chainiac:_2017} combines the ideas of skiplists and blockchains to realize \emph{skipchains}, an authenticated log with both back and forward long-distance links to implement a distributed authenticated software-release ledger. Chainiac relies on digital collective signatures to implement forward links, which are not available in permissionless Proof-of-Work chains such as Bitcoin. Long distance links are particularly well adapted to navigate a well-identified subset of a blockchain (such as a package's releases). They were not however directly designed to handle the kind of dependencies captured by the UTXO model.

An entirely different approach to scaling blockchains for lightweight nodes is the use of Non-Interactive Proofs of Proof-of-Work (NIPoPoWs)~\cite{kiayias_non-interactive_2017} that enable constant size queries.
NIPoPoWs strive for minimal cost of proof of inclusion of a transaction in a chain, thus reducing to a minimum the bandwidth requirements of lightweight nodes.
NIPoPoWs however do not aim at offering improved security for lightweight nodes as we do with Dietcoin.

%% file: dietcoin_conclusion.tex
\section{Conclusion}
\label{sec:conclusions}

In this report, we have presented the design of Dietcoin, that proposes a new form of Bitcoin nodes that strengthens the security guarantees of lightweight SPV nodes by bringing them closer to those of full Bitcoin nodes. The Dietcoin protocol enables low-resource nodes to verify the transactions contained in blocks without constructing a full-fledged UTXO set.
In our protocol, diet nodes download from full nodes parts of the UTXO set they need in order to verify a block, or a subchain of blocks, of interest. Diet nodes are able to detect any tampering of the UTXO set itself, at a cost that remains affordable for low-resource devices, both in terms of communication and computing overhead. In our approach, Dietcoin-enabled full nodes split their UTXO set into small shards,
and enable diet nodes to download only the shards that are relevant to the transactions in the block, while verifying that these shards do indeed corresponds to the state of the UTXO set for the block they are verifying.

% In the future, we plan to explore other sharding policies to optimize performances, as well as the ability for client to cache the UTXO shards they download.

%% file: algos/algo_tables.tex
% -*- tex-main-file: "dietcoin" compile-command: "make pdf" ispell-dictionary: "american" -*

\begin{table}[t]
%\begin{minipage}{.5\linewidth}
	\centering
	\begin{tabular}{|l|l|}
		\hline
		$\bloomFilter(\lv{keys})$ & Compute a bloom filter from $\lv{keys}$ \\
		$\buildMRoot(\lv{hashes})$ & Compute the Merkle root from $\lv{hashes}$ \\
%		$\hashFunction(x)$ & $\sha{x}$ \\
%		$\difficulty(\lv{header})$ & Difficulty of $\hashFunction(\lv{header})$ \\
%		$\depth(\lv{header})$ & Index of $\lv{header}$ starting from the tip \\
		$\getShardKey(\lv{txId})$ & Apply the sharding algorithm to $\lv{txId}$ \\
		$\height(\lv{header})$ & Index of $\lv{header}$ starting from the genesis block \\
%		$\getShardingPolicy(\lv{block})$ & Get the UTXO sharding policy from $\lv{block}$ \\
%		$\initTemplate()$ & Pre-fill the block header and the coinbase \\
%		$\updateBestTip()$ & Recompute the best tip in $\lv{headerStore}$ \\
%		$\updateDifficulty()$ & Adjust the difficulty of finding a block \\
        $\updateMTreeInPlace(\lv{MTree}, \lv{dataset})$ & Update the hashes of $\lv{MTree}$ with the new value of $\lv{dataset}$ \\
		$\mathrm{verifyHeaders}(\lv{headers})$ & Add $\lv{headers}$ to the chain, return the hash of the new chain tip \\
		\hline
	\end{tabular}
	%\caption{Descriptions of non explicitly defined functions.}
%\end{minipage}
%\begin{minipage}{.5\linewidth}
%	\centering
%	\begin{tabular}{|l|l|}
%		\hline
%		$\gv{blockStore}$ & Dictionary of blocks \\
%		$\gv{headerStore}$ & Tree of stored block headers \\
%		$\gv{shards}$ & Dictionary of UTXO shards \\
%		$\gv{targetDifficulty}$ & Minimum difficulty to validate a header \\
%		$\gv{tipId}$ & Hash of the most recent block (tip) in the chain \\

%		$\gv{utxoTree}$ & Merkle tree of UTXO shards \\
%		\hline
%	\end{tabular}
%	\caption{Descriptions of global variables for a node $p$.}
%\end{minipage}
\end{table}

%%% Local Variables:
%%% mode: latex
%%% End:

%% file: algos/algo_spv_compact.tex
% -*- tex-main-file: "dietcoin" compile-command: "make pdf" ispell-dictionary: "american" -*

\begin{figure}[t]
	\commentPLR{Fix indentation messed up by the bullets}

	\commentPLR{Add text to clarify notation? () list, \{\} object}
	\begin{algorithm}[H]
		\caption{-- SPV [$\circ$] and Diet [$\circ\bullet$] node $p$ (compacted)}
		\label{algo:spv_compact}
		\small

		\begin{algorithmic}[1]
			\State $\kw{parameters:}~\circ\gv{pubKeys}, \bullet~\gv{maxDepth}, \bullet~\gv{maxLength}$
			\State $\kw{global variables:}~\circ\gv{headerStore}, \circ~\gv{tipId}, \bullet~\gv{highestVerified}$
			\smallskip

\bc			\Procedure{\fn{updateChain}}{~} \Comment{App interface}
				\State $\lv{filter} \leftarrow \bloomFilter(\gv{pubKeys})$
				\State $(\{\lv{header}, \lv{txMTree}, \lv{txs}\}) \leftarrow \kw{send}~\fn{queryMerkleBlocks}(\gv{tipId}, \lv{filter})$
				\State $\gv{tipId} \leftarrow \mathrm{verifyHeaders}((\lv{header}))$

				\ForAll{$\{\lv{header}, \lv{txMTree}, \lv{txs}\} \in (\{\lv{header}, \lv{txMTree}, \lv{txs}\})$}
%					\LeftComment{Ignore false positives inherent to bloom filter usage}
					\If{$\forall k \in \gv{pubKeys}: k \notin \{\lv{tx.inputs} \cup \lv{tx.outputs}\}$}
						\State $\kw{continue}$ \Comment{Ignore bloom filter false positives}
					\EndIf
					\State $\kw{assert}(\forall \lv{tx} \in \lv{txs}: \hashFunction(\lv{tx}) \in \lv{txMTree})$
					\State $\lv{builtTxMRoot} \leftarrow \buildMRoot(\lv{txMTree})$
					\State $\kw{assert}(\lv{builtTxMRoot} = \lv{header.txMRoot}$)
\dc					\State $\fn{verifyBlocksUpTo}(\height(\lv{header}))$
\bc					\State $\kw{callback}(\lv{txs}, \lv{header}$) \Comment{Callback to app}
				\EndFor
			\EndProcedure
			\smallskip

\dc			\Procedure{\fn{verifyBlocksUpTo}}{$\lv{last}$}
				\LeftComment{Do not verify blocks twice or below $\gv{maxDepth}$}
				\State $\lv{first} \leftarrow \kw{max}(\gv{highestVerified}, \height(\gv{tipId}) - \gv{maxDepth})$
				\LeftComment{Verify up to $\gv{maxLength}$ blocks}
				\State $\lv{first} \leftarrow \kw{max}(\lv{first}, \lv{last} - \gv{maxLength})$

%				\LeftComment{Fallback to SPV mode if full verification is disabled}
				\If{$\lv{first} \geq \lv{last}$}
					\State $\kw{return}$ \Comment{Fallback to SPV mode}
				\EndIf
%				\State $\lv{last} \leftarrow \height(\lv{header})$
				\LeftComment{The first UTXO Merkle root is not verified}
				\State $\lv{utxoMRoot} \leftarrow \kw{send}~\fn{queryUtxoMRoot}(\hashFunction(\gv{headerStore}[\lv{first}]))$
				\ForAll{$\lv{blockId}$ of height $\in [\lv{first} + 1, \lv{last}]$}
					\State $\lv{block} \leftarrow \kw{send}~\fn{queryBlock}(\lv{blockId})$
%					\LeftComment{Download and check the authenticity of the UTXO shards to verify the block}
					\State $\{\lv{shards}, \lv{utxoMTree}\} \leftarrow \kw{send}~\fn{queryUtxos}(\lv{blockId})$
					\State $\kw{assert}(\forall \lv{shard} \in \lv{shards}: \hashFunction(\lv{shard}) \in \lv{utxoMTree})$
					\State $\lv{builtUtxoMRoot} \leftarrow \buildMRoot(\lv{utxoMTree})$
					\State $\kw{assert}(\lv{builtUtxoMRoot} = \lv{utxoMRoot})$
%					\LeftComment{Verify transactions and update UTXO shards}
					\ForAll{$\lv{tx} \in \lv{block.transactions}$}
						\ForAll{$\lv{i} \in \lv{tx.inputs}$}
							\State $\lv{shard} \leftarrow \lv{shards}[\getShardKey(\lv{i})]$
							\State $\kw{assert}(\lv{i} \in \lv{shard} \wedge \text{valid proof of ownership of } \lv{i})$
							\State $\lv{shard}.\mathrm{remove}(\lv{i})$
						\EndFor
						\ForAll{$\lv{o} \in \lv{tx.outputs}$}
							\State $\lv{shards}[\getShardKey(\lv{o})].\mathrm{add}(\lv{o})$
						\EndFor
					\EndFor
%					\LeftComment{Update and verify the UTXO Merkle root in the block}
					\State $\lv{utxoMTree} \leftarrow \updateMTreeInPlace(\lv{utxoMTree}, \lv{shards})$
					\State $\lv{utxoMRoot} \leftarrow \buildMRoot(\lv{utxoMTree})$
					\State $\kw{assert}(\lv{utxoMRoot} = \lv{block}.\lv{utxoMRoot})$
%					\State $\kw{callback}(\lv{txs}, \lv{header})$ \Comment{Callback to app}
					\State $\gv{highestVerified} \leftarrow \height(\lv{blockId})$
				\EndFor
			\EndProcedure

		\end{algorithmic}
	\end{algorithm}
\end{figure}

%%% Local Variables:
%%% mode: latex
%%% End: